\documentclass[10pt, aps,prl,twocolumn,showpacs,superscriptaddress]{revtex4-2}

\usepackage[colorlinks=true,linkcolor=blue,citecolor=blue,urlcolor=magenta]{hyperref}

\usepackage{amsfonts,mathrsfs,amsmath,amsthm,amssymb}
\usepackage{epsfig}
\usepackage{bm}
\usepackage{array,multirow}
\usepackage{booktabs}
\usepackage{graphicx}
\usepackage{upgreek}
\usepackage{ulem}

\newcommand{\inlinesection}[1]{\par\smallskip\noindent\textit{#1}.}

\begin{document}  
\title {\bf 
Optimized quantum sensor networks for ultralight dark matter detection}

\author{Adriel I. Santoso} 
\affiliation{Department of Mechanical and Aerospace Engineering, Tohoku University, Sendai 980-0845, Japan}

\author{Le Bin Ho} 
\thanks{Electronic address: binho@fris.tohoku.ac.jp}
\affiliation{Frontier Research Institute 
for Interdisciplinary Sciences, 
Tohoku University, Sendai 980-8578, Japan}
\affiliation{Department of Applied Physics, 
Graduate School of Engineering, 
Tohoku University, 
Sendai 980-8579, Japan}

\date{\today}

\begin{abstract}
Dark matter (DM) remains one of the most compelling unresolved problems in fundamental physics, motivating the search for new detection approaches. We propose a network-based quantum sensor architecture to enhance sensitivity to ultralight DM fields. Each node in the network is a superconducting qubit, interconnected via controlled-Z gates in symmetric topologies such as line, ring, star, and fully connected graphs. We investigate four- and nine-qubit systems, optimizing both state preparation and measurement using a variational quantum metrology framework. This approach minimizes the quantum and classical Cram{\'e}r-Rao bounds to identify optimal configurations. Bayesian inference is employed to extract the DM-induced phase shift from measurement outcomes. Our results show that optimized network configurations significantly outperform conventional GHZ-based protocols while maintaining shallow circuit depths compatible with noisy intermediate-scale quantum hardware. Sensitivity remains robust under local dephasing noise.
These findings highlight the importance of network structure in quantum sensing and point toward scalable strategies for quantum-enhanced DM detection.
\end{abstract}
%
%
\maketitle

\inlinesection{Introduction}
Dark matter (DM), comprising about 27\% of the universe’s mass-energy content, remains a central mystery in modern physics \cite{RevModPhys.90.045002, PhysRevD.98.030001}. It is essential for explaining galactic binding and the formation of large-scale cosmic structures \cite{10.1093/mnras/183.3.341, annurev:/content/journals/10.1146/annurev.astro.36.1.599}. 
The idea was first suggested by Zwicky’s discovery of missing mass in galaxy clusters \cite{Zwicky1933} and later supported by Rubin’s measurements of galactic rotation curves \cite{Rubin1980}.
Despite extensive searches for candidates such as WIMPs, axions, and sterile neutrinos, none have been confirmed \cite{annurev:/content/journals/10.1146/annurev-astro-082708-101659,Baudis_2016,BOYARSKY2012136,peter2012darkmatterbriefreview,PhysRevD.94.095010,Schumann_2019,Roszkowski_2018}. Detecting weakly interacting or light candidates like axions or hidden photons remains difficult due to their extremely weak couplings with standard matter \cite{ringwald2012exploringroleaxionswisps}. 
Revealing the fundamental nature of DM is still a central goal of astrophysics and particle physics.
 
Recent advances in quantum technologies offer new opportunities for DM detection. Superconducting qubits provide exceptional sensitivity to weak perturbations owing to their tunability and long coherence times~\cite{PRXQuantum.4.020101,PhysRevLett.131.211001}. Single qubits have been proposed to detect ultralight DM via shifts in electromagnetic fields or spin orientations~\cite{PhysRevLett.126.141302,10409573}, while entangled qubits can further enhance sensitivity~\cite{PhysRevLett.133.021801,PhysRevD.110.115021}. Quantum metrology, exploiting correlations beyond classical limits, has been employed to analyze detection precision~\cite{PhysRevA.111.012601}.

Quantum sensor networks, built from interconnected sensors, enable enhanced sensitivity and robustness by distributing correlations across nodes~\cite{PhysRevLett.120.080501,Dailey2021,PhysRevLett.124.110502,Nguyen2024,Le2023}. However, their performance strongly depends on topology, e.g., poorly structured networks may degrade sensing. Network optimization is thus crucial for maximizing quantum sensing performance.

In this Letter, we propose a quantum sensor network for enhanced DM detection, where each node is a superconducting qubit linked by controlled-Z (CZ) gates. We study symmetric topologies, i.e., line, ring, star, and fully connected. We analyze four- and nine-qubit systems to assess their sensing performance.

To optimize sensing performance, we use variational quantum metrology (VQM)~\cite{Le2023,Meyer2021,MacLellan2024}, which minimizes the quantum and classical Cram{\'e}r-Rao bounds (QB, CB) by updating states and measurements. The DM-induced phase is then estimated via Bayesian inference.

Our results show that optimized quantum sensor networks surpass GHZ-based protocols, achieving high precision with shallow circuits suitable for noisy intermediate-scale quantum (NISQ) devices. The precision scales favorably with qubit number and remains robust against dephasing noise.

Beyond DM detection, our optimization methods and network architectures extend to quantum sensing applications such as gravitational waves, spectroscopy, and magnetometry~\cite{DeMille2024}, providing a practical framework for tackling major scientific challenges.

\inlinesection{DM detection via superconducting qubits}
Recent studies have used transmon qubits as quantum sensors for detecting ultralight DM~\cite{PhysRevLett.131.211001, PhysRevLett.126.141302, PhysRevLett.133.021801, PhysRevD.110.115021,10409573}. In these schemes, a coherently oscillating DM field induces an effective electric field, $\bar{E}^{(\text{eff})}$, which drives Rabi-like transitions between the ground and excited states of the qubit. The interaction is described by the effective Hamiltonian
\begin{equation}
\mathcal{H} = \omega |e\rangle\langle e| - 2\eta \cos\big(m_{\text{DM}}t - \alpha\big)\big(|e\rangle\langle g| + |g\rangle\langle e|\big),
\end{equation}
where $\omega$ is the qubit energy splitting, $m_{\text{DM}}$ is the DM mass, $\alpha$ is the DM random phase offset, and $|g\rangle$, $|e\rangle$ denote the ground and excited states, respectively. The coupling strength is given by
\(
\eta \equiv \frac{\sqrt{C\omega}}{2\sqrt{2}} \bar{E}^{(\text{eff})} d \cos\Theta,
\)
where $C$ is the transmon’s capacitance, $d$ is the distance between two plates, obeying $C \propto 1/d$, and $\Theta$ is the angle between $\vec{E}^{(\text{eff})}$ and the normal vector of the capacitor plates. 
Increasing $d$ decreases $C$ and increases the coupling $\eta$. For transmon qubits, $d \sim 10\text{–}100\,\mu\text{m}$. Fabrication limits, such as lithographic resolution, alignment, and chip size, further restrict $d$: too small $d$ is prone to defects, while very large $d$ increases chip area and parasitic effects.
In simulations, $\cos^2\Theta \approx 1/3$ to account for angular fluctuations within the DM coherence time~\cite{PhysRevLett.133.021801, PhysRevD.110.115021}.

The evolution is governed by the Schr\"odinger equation, 
\(
i\frac{d}{dt}|\Psi(t)\rangle = \mathcal{H}|\Psi(t)\rangle,
\)
where $|\Psi(t)\rangle = \psi_g(t)|g\rangle + \psi_e(t)|e\rangle$. Within the coherence time $\tau \simeq \min(\tau_{\text{DM}}, \tau_q)$, where $\tau_{\text{DM}}$ and $\tau_q$ are the coherence times of the DM field and the qubit, respectively, the evolution is described by a unitary operator $U_{\text{DM}}(\tau)$
\begin{equation}
\begin{pmatrix}
\psi_g(\tau)\\
\psi_e(\tau)
\end{pmatrix}
= U_{\text{DM}}(\tau)
\begin{pmatrix}
\psi_g(0)\\
\psi_e(0)
\end{pmatrix}.
\end{equation}

In the resonant regime, $\omega = m_{\text{DM}}$, the interaction simplifies under the rotating wave approximation, yielding
\begin{align}\label{eq:Udm}
U_{\text{DM}}(\tau) =
\begin{pmatrix}
\cos\delta & i e^{-i\alpha} \sin\delta \\
i e^{i\alpha} \sin\delta & \cos\delta
\end{pmatrix}, \quad \delta \equiv \eta \tau,
\end{align}
where
\(
|\psi_\pm\rangle = \frac{1}{\sqrt{2}}\left(|g\rangle \pm e^{i\alpha}|e\rangle\right).
\)
When $\omega \neq m_{\text{DM}}$, the sensitivity decreases over a mismatch bandwidth $\Delta f \sim \tau^{-1}$.
For a fixed $\omega$, high sensitivity is maintained for $m_{\text{DM}} \in \omega \pm \tau^{-1}$, and $U_{\text{DM}}$ remains applicable \cite{PhysRevD.110.115021}.

In the individual-qubit scheme, each transmon acts as an independent sensor, giving a total excitation probability
\(
p_{g \rightarrow e}(\tau) \simeq N \delta^2\), for \(N\delta^2 \ll 1,\)
where $N$ the number of qubits.

To surpass this shot-noise limit, entangled protocols using GHZ states have been proposed~\cite{PhysRevLett.133.021801, PhysRevD.110.115021}. 
At $t_1$, a GHZ state is prepared
\begin{align}\label{eq:ghz}
|\Psi(t_1)\rangle = \frac{1}{\sqrt{2}} \Big( |g\rangle^{\otimes N} + |e\rangle^{\otimes N} \Big).
\end{align}
During the interaction time $\tau = t_2 - t_1$, each qubit acquires a phase $\delta$, giving
\begin{equation}
|\Psi(t_2)\rangle = \frac{1}{\sqrt{2}} \Big( e^{i N \delta} |+\rangle^{\otimes N} + e^{-i N \delta} |-\rangle^{\otimes N} \Big),
\end{equation}
where $|\pm\rangle = \frac{1}{\sqrt{2}}(|g\rangle \pm |e\rangle)$.

After interaction, a reversed GHZ operation maps the accumulated phase onto one qubit\begin{align}
|\Psi(t_f)\rangle = \Big( \cos(N\delta)\, |g\rangle + i \sin(N\delta)\, |e\rangle \Big) \otimes |+\rangle^{\otimes (N - 1)}.
\end{align}
Here, we assumed $t_f = t_2$, neglecting further evolution beyond the sensing interval. The entangled state accumulates phase as $N^2\delta^2$, giving quadratic sensitivity in $N$, in contrast to the linear scaling of individual-qubit strategies. This quadratic enhancement highlights a central advantage of entangled probes in quantum metrology, enabling parameter estimation beyond the shot-noise limit.

However, the optimal state for enhancing metrological precision must be a superposition of the minimum and maximum eigenstates of \( U_{\rm DM} \)~\cite{Giovannetti2011}, i.e, 
\(
|\Psi(t_1)\rangle = \frac{1}{\sqrt{2}}(|\psi_+\rangle^{\otimes N} + |\psi_-\rangle^{\otimes N}).
\)
In contrast, the conventional GHZ state in Eq.~\eqref{eq:ghz} does not yield improved precision [see Fig.~2(a)]. To overcome this limitation, we employ a network-based quantum sensor and optimize performance using VQM.

\begin{figure*}[t]
    \centering
    \includegraphics[width=0.8\linewidth]{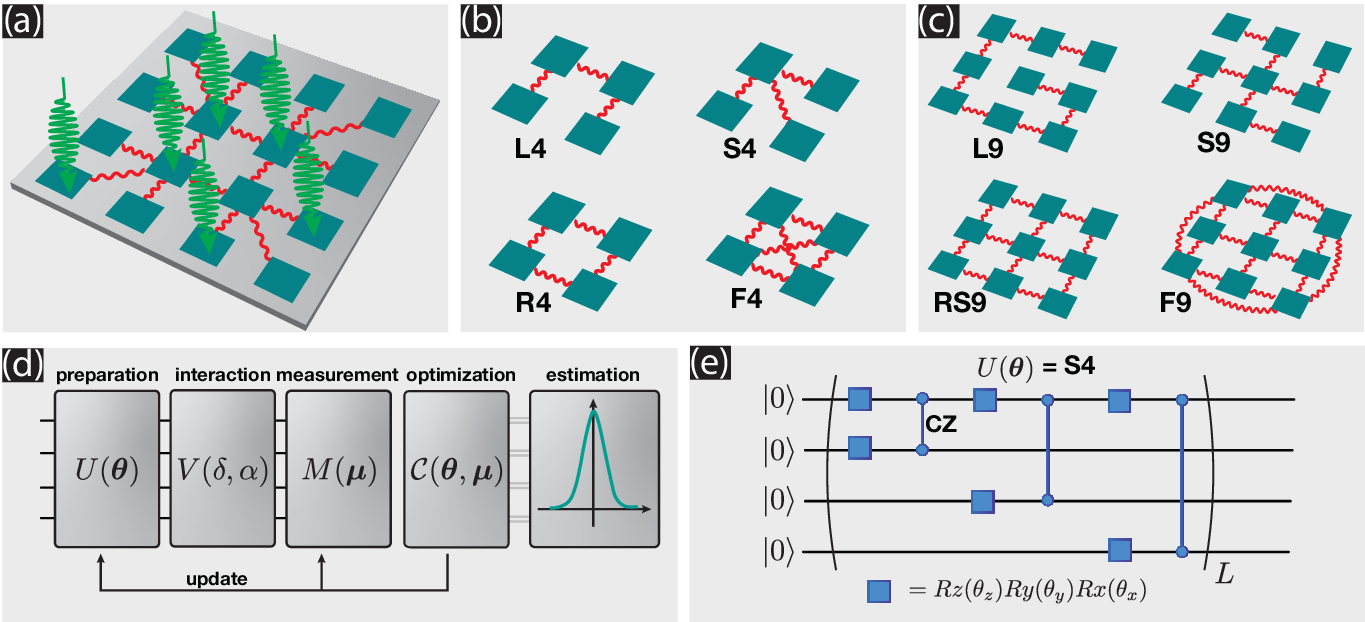}
    \caption{ (a) Sensor-network for DM detection: 2D superconducting qubit array with CZ gates forming a tunable network. (b)–(c) Configurations for $N=4$ and $N=9$. (d) VQM framework where $U(\bm\theta)$ and measurement $M(\bm\mu)$ come from (b, c); external evolution $V(\delta,\alpha)$ comes from the DM field; the cost $\mathcal{C}(\bm\theta,\bm\mu)$ optimized via QB and CB; reconstruction by Bayesian estimation. (e) Example variational circuit for S4.
    }
    \label{fig:1}
\end{figure*}

\inlinesection{Quantum-enhanced metrology viewpoint}
The precision in estimating $\delta$ is bounded by Cram\'er-Rao bounds
\begin{align}\label{eq:crb}
\nu\,\text{Var}(\delta) \geq \frac{1}{F(\delta)} \geq \frac{1}{Q(\delta)},
\end{align}
where \( \nu \) is the number of measurements, \( F(\delta) \) is the classical Fisher information (CFI), and \( Q(\delta) \) is the quantum Fisher information (QFI). 

The CFI, determined by the conditional probability \( p(m|\delta) \), is
\begin{align}
F(\delta) = \sum_m \frac{1}{p(m|\delta)} \left( \frac{\partial p(m|\delta)}{\partial \delta} \right)^2,
\end{align}
quantifying the information about $\delta$ from a given measurement strategy.

The QFI gives the maximum information attainable from all quantum measurements and depends only on the state $\rho_\delta$
\begin{equation}
Q(\delta) = \text{Tr}\left[\rho_\delta L_\delta^2\right],
\end{equation}
where \( \rho_\delta = |\Psi(t_2)\rangle \langle \Psi(t_2)| \) and \( L_\delta \) is the symmetric logarithmic derivative defined by 
\(2\partial\rho_\delta/\partial\delta  = L_\delta \rho_\delta + \rho_\delta L_\delta\). Larger QFI implies higher sensitivity and more precise estimation.

For non-entangled states, 
\(
Q(\delta) \propto N,
\) 
corresponding to the shot-noise limit, which represents the fundamental precision bound for classical strategies. For an entangled GHZ state, 
\(
Q(\delta) \propto 4N^2,
\)
demonstrating Heisenberg limit. 
However, as noted above, standard GHZ states do not always yield optimal precision. This motivates further exploration of adaptive protocols and optimal measurement strategies, as discussed below.

\inlinesection{Quantum networks for DM detection}
We model a network-based quantum sensor as a 2D array of superconducting qubits, as illustrated in Fig.~\ref{fig:1}(a). Each qubit is a node, and edges represent physical connections between nodes. Due to hardware constraints, each qubit is limited to a maximum of four connections, consistent with the restricted connectivity in current superconducting architectures. 
We examine systems of four and nine qubits, shown in Fig.~\ref{fig:1}(b) and Fig.~\ref{fig:1}(c), respectively. For the 4-qubit case, we analyze linear (L4), star (S4), ring (R4), and fully connected (F4) configurations. For the 9-qubit system, we study simular structures with linear (L9), star (S9), ring-star hybrid (RS9), and fully connected network (F9). 

\inlinesection{Variational quantum metrology}
VQM is a hybrid quantum-classical framework for optimizing quantum sensing protocols using parameterized quantum circuits~\cite{Le2023, Meyer2021, MacLellan2024}. 
In this scheme, the state is prepared by an ansatz $U(\bm{\theta})$, evolves under an external field such as $U_{\rm DM}$, and is then measured through an ansatz $M(\bm{\mu})$. The parameters $\bm{\theta}$ and $\bm{\mu}$ are variationally updated to minimize cost functions such as the Cram\'er-Rao bounds \cite{Le2023}. This hybrid framework enables adaptive sensing strategies, has been demonstrated on several platforms~\cite{Cimini2024,Nielsen2025}, and is well suited to near-term devices.

VQM shares a structure with other variational algorithms like variational quantum eigensolver (VQE) \cite{Peruzzo2014}, where a parameterized quantum circuit is optimized via classical feedback. While VQE minimizes energy to find ground states, VQM minimizes the Cram\'er-Rao bounds to improve the parameter estimation, highlighting their similar workflows.

\begin{figure*}[t]
    \centering
    \includegraphics[width=0.65\linewidth]{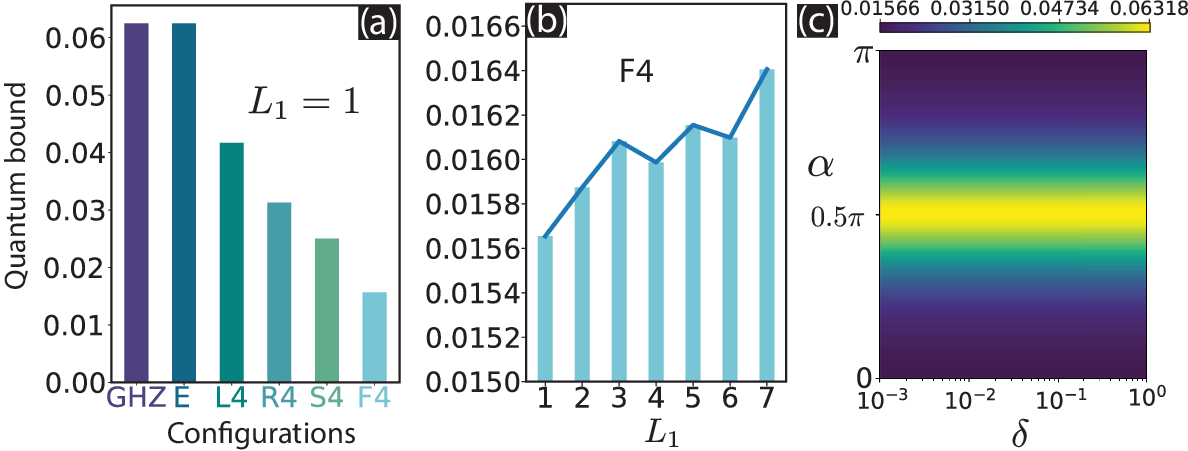}
    \caption{
(a) QBs for four-qubit system at $L_1=1$; optimized ansatze, especially F4, outperform GHZ and excited states. (b) QBs of F4 versus depth $L_1$; best precision at shallow $L_1=1$, showing efficiency. (c) Robustness of F4 at $L_1=1$; QB nearly constant with $\delta$ and periodic in $\alpha$ without re-optimization.
    }
    \label{fig:2}
\end{figure*}

We implement the VQM illustrated in Fig.~\ref{fig:1}(d), where \( U(\bm{\theta}) \) and \( M(\bm{\mu}) \) are structured according to the network in Figs.~\ref{fig:1}(b, c). These ansatze consist of rotation blocks \( U_{\rm rot} = R_z(\theta_z) R_y(\theta_y) R_x(\theta_x) \), inserted between each CZ gate~\cite{Le2023}. An example of \( U(\bm{\theta}) \) for S4 is provided in Fig.~\ref{fig:1}(e). We denote \( L_1 \) and \( L_2 \) as the number of layers in \( U(\bm{\theta}) \) and \( M(\bm{\mu}) \), respectively.
These ansatze use a linear number of parameters, keeping the optimization cost low for small systems. In larger Hilbert spaces, shallow circuits may lack expressivity \cite{Meyer2021,MacLellan2024}, whereas deeper circuits improve it but introduce risks such as barren plateaus \cite{Cerezo2021}, initialization sensitivity, and increased noise susceptibility. Therefore, selecting the optimal depth involves balancing expressivity, trainability, and robustness. 

The sensing process is described by the global unitary operator
\begin{align}
V(\delta, \alpha) = \bigotimes_{i=1}^N U_{\rm DM}(\delta, \alpha),
\end{align}
which models interactions between each qubit and the external DM field. Here, \( U_{\rm DM}(\delta, \alpha) \) is the single-qubit evolution operator defined in Eq.~\eqref{eq:Udm}.

\inlinesection{Optimization and estimation}
We first optimize \( \bm{\theta} \) by minimizing a cost function defined by \(
    \mathcal{C}(\bm\theta) = Q^{-1}(\delta),
\)
yielding the optimal configuration \( \bm{\theta}^* = \arg\min_{\{\bm{\theta}\}} \mathcal{C}(\bm{\theta}) \). Once \( \bm{\theta}^* \) is fixed, we optimize \( \bm{\mu}\) using a similar procedure, with the cost function is \(
    \mathcal{C}(\bm\mu) = F^{-1}(\delta).
\)
To update parameters, we use Adam optimizer. 

After optimizing the network state and measurement configuration, we apply Bayesian estimation to recover \( \delta \) from the measurement outcomes. Let \( m \in \{00\cdots0, \ldots, 11\cdots1\} \) denote an outcome in the computational basis, and \( p(\delta) \) the prior probability distribution representing initial knowledge of \( \delta \).  The likelihood function \( p(m|\delta) \) is given by the probability of observing outcome \( m \) for a given \( \delta \), computed from the final quantum state.

Using Bayes' theorem, the posterior distribution \( p(\delta | m) \) is given by
\begin{align}
    p(\delta | m) = \frac{p(m | \delta) p(\delta)}{p(m)},
\end{align}
where \( p(m) = \int p(m | \delta) p(\delta) \, d\delta \) is the marginal likelihood.
We choose $p(\delta)$ as the probability density function (pdf). This choice has minimal impact on the final results because each measurement updates the posterior, which then serves as the new prior. For a large number of measurements, this sequential updating ensures the final estimate is largely insensitive to the initial prior.

The posterior distribution \( p(\delta | m) \) encodes all information about \( \delta \) after the measurement. The estimate \( \bar{\delta} \) is given by the posterior mean, and the precision is quantified by the variance \( \mathrm{Var}(\delta) \).

\inlinesection{Numerical results}
For VQM optimization, we fix \( \delta = 0.05 \) and \( \alpha = 0 \). The parameters \( \bm{\theta} \) and \( \bm{\mu} \) are initialized randomly. Measurement outcomes are simulated using the Primitives sampler from Qiskit.

\inlinesection{Quantum bound} 
We evaluate the performance of various ansatze for a four-qubit system (\( N = 4 \)) and present the results in Fig.~\ref{fig:2}(a) for \( L_1 = 1 \). The GHZ and excited state (E) do not achieve QBs comparable to those of the optimized configurations. As noted earlier, the GHZ state is suboptimal under evolution by \( U_{\rm DM} \) because it is not aligned with the eigenstates associated with the eigenvalues of the unitary. Consequently, its performance is comparable to that of the excited state.

By contrast, the optimized ansatze, particularly those with more connections such as L4, R4, S4, and especially F4, outperform the others. 
The best performance of F4 is attributed to its higher connectivity and larger number of trainable parameters, which significantly enhance quantum state correlations and contribute to greater circuit expressiveness.

In Fig.~\ref{fig:2}(b), we explore the QB of F4 for different layers \( L_1 \). Remarkably, the best performance is achieved with \( L_1 = 1 \). The result indicates that high precision is achievable with shallow circuits, highlighting the method’s efficiency and practicality for NISQ devices.
Nonetheless, practical limitations, including gate errors ($\sim$ 0.1–1\%), finite coherence times, and restricted connectivity in superconducting devices, can affect performance. A detailed analysis will be addressed in future work.

With the fixed F4, \( L_1 = 1 \) configuration, 
we next examine the QB as a function of \( \delta \) and \( \alpha \), as shown in Fig.~\ref{fig:2}(c). For a fixed \( \alpha \), the QB remains constant over a wide range of \( \delta \), from \( 10^{-3} \) to \( 10^0 \). This robustness implies that re-optimization is not required for each new value of \( \delta \), offering significant practical advantages for sensing the DM field. When varying \( \alpha \), the QB shows a periodic behavior, peaking at \( \alpha = \pi/2 \) and returning to its original value at \( \alpha = 0 \) and \( \pi \). While the optimization was performed at $\alpha = 0$, we verified that the protocol performs robustly for all $\alpha$ values. The QB shows only an increase around $\alpha = \pi/2$, which does not significantly affect the overall measurement precision.

These findings highlight the potential of our network-based approach to generate high-performance quantum resources for metrology and sensing. Next, we optimize CB.

\inlinesection{Classical bound}
To achieve the best CB, it is essential to use the optimal measurement operators, represented as positive operator-valued measures (POVMs). In our VQM framework, these measurement operators are optimized through the ansatz \( M(\boldsymbol{\mu}) \). Specifically, the circuit for \( M(\boldsymbol{\mu}) \) is chosen to be the inverse of the input circuit, i.e., \( {\rm F}4^\dagger \), with a variable number of layers \( L_2 \).

\begin{figure}
    \centering
    \includegraphics[width=\columnwidth]{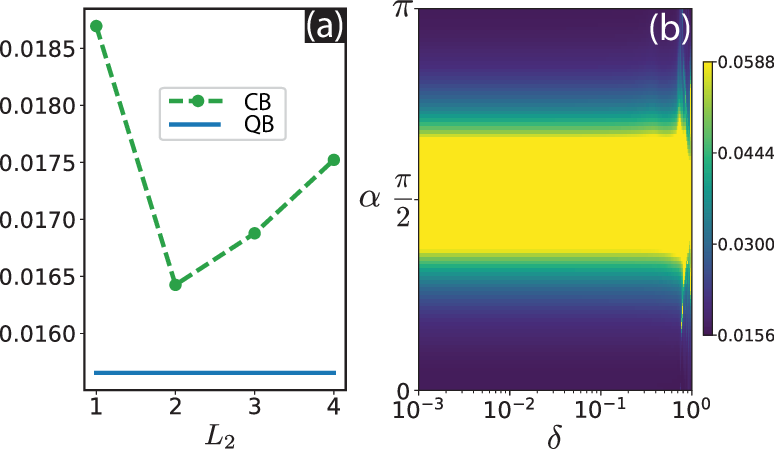}
    \caption{ 
(a) CB versus depth $L_2$; increasing $L_2$ decreases CB toward the QB, then increases. (b) CB versus $\delta$ and $\alpha$; optimized measurements keep CB nearly constant, with slight growth at large $\delta$ and periodic variation with $\alpha$, consistent with QBs.
}
    \label{fig:3}
\end{figure}

Figure~\ref{fig:3}(a) shows that with a shallow measurement circuit (\( L_2 = 1 \)), the CB is larger than the QB, indicating poor measurement performance. As \( L_2 \) increases, the CB approaches the QB, reflecting improved accuracy. However, beyond an optimal depth, further increasing \( L_2 \) will raising the CB and reducing precision. This non-monotonic behavior suggests that deeper circuits may suffer from limited expressibility, overparameterization, or barren plateaus, where the optimization landscape becomes flat. The gap between CB and QB arises from the inability of the measurement ansatz to fully approximate the optimal POVM associated with the QFI.

In Fig.~\ref{fig:3}(b), we examine how the CB varies with \(\delta\) and \(\alpha\). Our optimized measurement configuration yields a constant, minimal CB over a wide range of \(\delta\), similar to the behavior observed for the QB in Fig.~\ref{fig:2}(c). However, when \(\delta\) becomes relatively large (on the order of \(10^0\)), the CB begins to increase. This rise does not significantly impact our conclusions, as the physically relevant regime for the DM field corresponds to small \(\delta\) values. 

Regarding \( \alpha \), we observe periodic variations in the CB, consistent with the behavior of the QB. This periodicity indicates that estimation performance depends on \( \alpha \), but the optimal configuration remains robust across its range.

\begin{figure}[t]
    \centering
    \includegraphics[width=\linewidth]{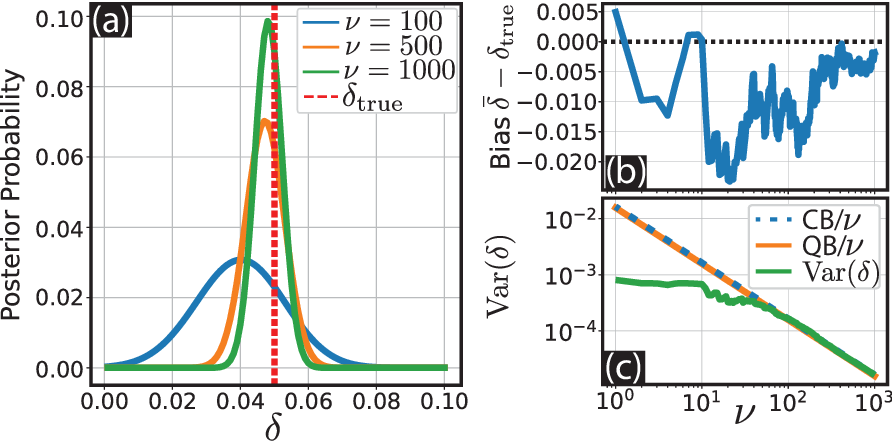}
    \caption{(a) Posterior versus phase $\delta$ for different $\nu$; the peak approaches the true value as $\nu$ grows. (b) Bias $\bar\delta-\delta_{\rm true}$ converges to zero with $\nu$. (c) Variance Var($\delta$) versus $\nu$, compared with ${\rm CB}/\nu$ and ${\rm QB}/\nu$, showing agreement and optimal scaling.
    }
    \label{fig:4}
\end{figure}

\inlinesection{Bayesian estimation} In Fig.~\ref{fig:4}(a), we plot the posterior probability distribution as a function of \(\delta\) for different numbers of measurements \(\nu\). As \(\nu\) increases, the peak shifts closer to the true value of \(\delta\), demonstrating improved estimation accuracy.

Figure~\ref{fig:4}(b) shows the bias $\bar\delta - \delta_{\rm true}$ versus $\nu$. The bias gradually approaches zero as $\nu$ increases, demonstrating that the estimator becomes asymptotically unbiased with more measurement data.

In Fig.~\ref{fig:4}(c), we plot Var($\delta$) against \(\nu\) and compare it with the theoretical predictions given by the QB and CB, both scaled as \(1/\nu\). The results show excellent agreement: the variance closely follows the trends of \({\rm QB}/\nu\) and \({\rm CB}/\nu\), confirming that our method achieves the optimal scaling expected from quantum estimation theory. For larger \(\nu\), the estimator becomes unbiased, and the bounds in Eq.~\eqref{eq:crb} hold.

\begin{figure}
    \centering
\includegraphics[width=\linewidth]{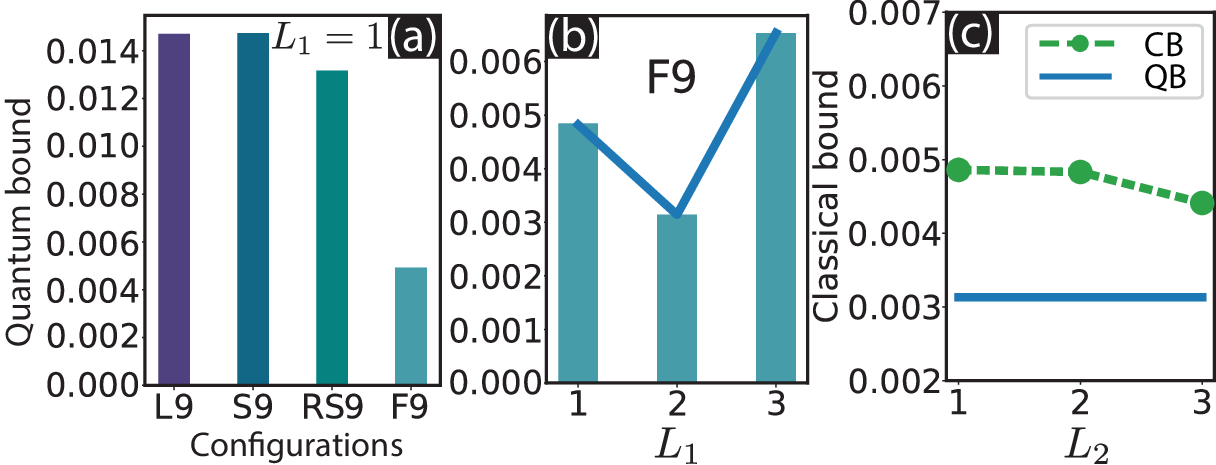}
    \caption{(a) QBs for nine qubits ($N=9$) using ansatze in Fig.~\ref{fig:1}(c); F9 gives the best QB, above the four-qubit case. (b) QBs versus~$L_1$ for F9. (c) CBs versus ~$L_2$ for F9.
    }
    \label{fig:5}
\end{figure}

\inlinesection{For $N=9$}
We evaluate the QB and present the results in Fig.~\ref{fig:5}. Using the same procedure as in Fig.~\ref{fig:2}(a) for the four-qubit system, we examine multiple configurations shown in Fig.~\ref{fig:1}(c). Among these, the fully connected F9 delivers the highest QB performance.

When compared to \( N = 4 \) , the QB for \( N = 9 \) is significantly improved. This improvement is expected because increasing the number of qubits generally enhances the achievable precision in quantum sensing, as predicted by quantum metrology theory.

In Fig.~\ref{fig:5}(b), we investigate the QB as a function of \( L_1 \). Different from four-qubit case, here, the best performance is obtained when (\( L_1 = 2 \)). This result suggests that even for larger systems, a shallow and well-optimized circuit is sufficient to achieve high-precision estimation. 

Figure~\ref{fig:5}(c) shows the CB versus $L_2$. As $L_2$ increases, the CB improves and approaches the QB. However, it does not reach the QB due to limitations in the measurement strategy. In this analysis, we evaluate up to $L_2 = 3$.
Remarkably, the precision in this case is better than that of the four-qubit case.

\inlinesection{Noisy effect}
We analyze the efficiency of a quantum sensor network under local dephasing noise. Each qubit undergoes dephasing described by the Kraus operators
\(
K_0 = \begin{pmatrix}
1 & 0 \\
0 & \sqrt{1 - \lambda}
\end{pmatrix},
K_1 = \begin{pmatrix}
0 & 0 \\
0 & \sqrt{\lambda}
\end{pmatrix}
\),
where $\lambda \in [0,1]$ characterizes the dephasing strength. The channel acts as
\(
    \mathcal{E}(\rho) = K_0 \rho K_0^\dagger + K_1 \rho K_1^\dagger.
\)

\begin{figure}
    \centering
\includegraphics[width=\linewidth]{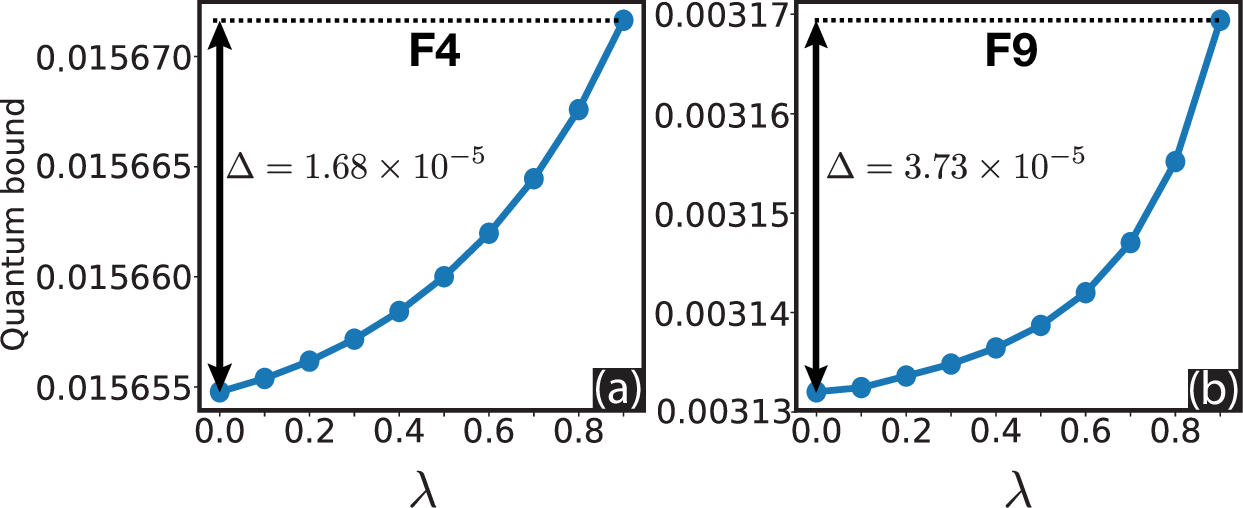}
    \caption{
   QBs under dephasing noise for (a) F4 and (b) F9. Sensitivity decreases slightly with increasing $\lambda$, but even at $\lambda=0.9$ the rise is small: $\Delta=1.68\times10^{-5}$ (a) and $\Delta=3.73\times10^{-5}$ (b).
    }
    \label{fig:6}
\end{figure}

The numerical results for the QB versus $\lambda$ are shown in Fig.~\ref{fig:6} for F4 (a) and F9 (b). No re-optimization is performed for each noise level; instead, the optimal configuration setup at $\lambda = 0$ is used throughout. Remarkably, the QBs slightly increase even as the noise strength rises to $\lambda = 0.9$. The difference between the QB at $\lambda = 0.9$ and the noiseless case is $\Delta = 1.68 \times 10^{-5}$ for F4 and $\Delta = 3.73 \times 10^{-5}$ for F9. This demonstrates that the optimal sensor networks remain robust under dephasing, with only minimal degradation in sensitivity even without re-optimization.

\inlinesection{Conclusion}
We proposed and demonstrated a network-based quantum sensing strategy for enhancing DM detection. By structuring superconducting qubits into networks with specific symmetries and optimizing their connections using VQM, we achieved substantial improvements in sensing precision compared to conventional GHZ-based protocols. Both the QB and CB were systematically minimized, and Bayesian estimation validated the enhanced sensitivity of the optimized networks. The optimal configurations maintained shallow circuit depths, making them feasible for implementation on NISQ devices. Our findings emphasize the significant impact of network design on quantum metrology performance and suggest that well-structured quantum network can serve as powerful platforms for DM searches and for a wide range of high-precision sensing applications. Future works may extend these results by exploring larger networks, adaptive strategies, and integration with error mitigation techniques to further push the limits of quantum-enhanced detection.

\inlinesection{Acknowledgments} 
This paper is supported by  
JSPS KAKENHI Grant Number 23K13025.

\inlinesection{Data availability} No data were created or analyzed in
this study.

\begin{figure}
    \centering
\includegraphics[width=0.8\linewidth]{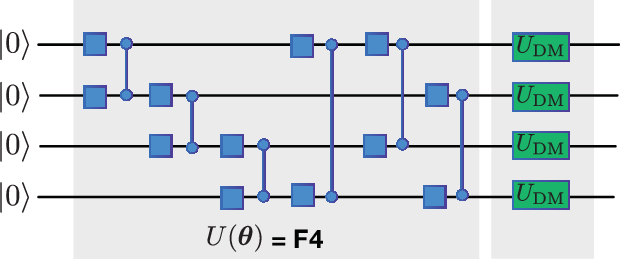}
    \caption{Quantum circuit for optimizing the QB in VQM.
    }
    \label{fig:7}
\end{figure}

\inlinesection{Appendix: Variational Quantum Metrology Optimization Method}
We detail the optimization process employed in variational quantum metrology (VQM). Consider the F4 configuration as an example, with the quantum circuit illustrated in Fig.\
~\ref{fig:7}. The system is initialized in the state $|\boldsymbol{0}\rangle$, which evolves to
\(
|\Psi(\delta, \alpha, \boldsymbol{\theta})\rangle = V(\delta, \alpha) U(\boldsymbol{\theta})|\boldsymbol{0}\rangle.
\)
The QFI is given by
\begin{align}
    Q = 4 \left( \langle \partial_\delta \Psi | \partial_\delta \Psi \rangle - \left| \langle \Psi | \partial_\delta \Psi \rangle \right|^2 \right),
\end{align}
where dependence on $\delta, \alpha, \boldsymbol{\theta}$ is omitted for brevity.

Consider the cost function as $\mathcal{C}(\boldsymbol{\theta}) = 1 / Q$, and minimize it using the Adam optimizer. The update rule at iteration $t$ is
\begin{align}
    \boldsymbol{\theta}_{t+1} = \boldsymbol{\theta}_t - \eta \frac{\hat{m}_t}{\sqrt{\hat{v}_t} + \epsilon},
\end{align}
where \(
m_t = \beta_1 m_{t-1} + (1 - \beta_1) g_t, \quad
v_t = \beta_2 v_{t-1} + (1 - \beta_2) g_t^2,
\hat{m}_t = \frac{m_t}{1 - \beta_1^t}, \quad
\hat{v}_t = \frac{v_t}{1 - \beta_2^t},
\)
and $g_t = \nabla_{\boldsymbol{\theta}} \mathcal{C}(\boldsymbol{\theta}_t)$. We use $\beta_1 = 0.9$, $\beta_2 = 0.999$, and $\epsilon = 10^{-8}$. The process is repeated until convergence of $\mathcal{C}(\boldsymbol{\theta})$.

To evaluate the required derivatives, i.e., $|\partial_\delta \Psi\rangle$ and $\nabla_{\boldsymbol{\theta}} \mathcal{C}(\boldsymbol{\theta})$, we employ the parameter-shift rule \cite{Ho2023}.

\bibliography{refs}

\end{document}